# Magnetically induced linear and nonreciprocal and tunable transparency

A.H.Gevorgyan

Far Eastern Federal University, 10 Ajax Bay, Russky Island, Vladivostok 690922, Russia; e-mall: agevorgyan@ysu.am

We report the discovery of a new effect, namely the effect of magnetically induced transparency. The effect is observed in magnetically active helically structured periodical medium. By changing the external magnetic field and absorption we can tune the frequency and linewidth of transparency band.

Electromagnetically induced transparency (EIT) is defined as coherent optical nonlinearity that makes a medium transparent in a narrow spectral range around the absorption line, which firstly was observed in a three-level atomic system and was caused by destructive interference between two pathways excited by two laser beams [1]. But due to experimental conditions such as extremely low temperature and high intensity laser, the research and potential application of electromagnetically induced transparency in atomic systems are seriously limited. Later, this concept was extended to the classical optical systems using gas-phase atomic [2,3], metamaterial/metasurface [4-7], plasmonic [8-9], optical [10-17], optomechanical [18-20] and superconducting [21,22] systems, which allow experimental implementation with incoherent light and operation at room temperature [23,24]. Among them, all-optical analogues of EIT realized in optical resonant systems, such as metamaterial, plasmonic, photonic crystal and whispering-gallery-mode resonators [17]. Thus, in recent years meta-atom (meta-system) based EIT analogs using coupled bright and dark resonators have been developed to mimic quantum destructive interference between excitation states in three-level atomic systems, resulting in a sharp transmission window within a broad absorption spectrum [8,9]. Such structures have ignited intensive research interests, as EIT produces enhanced transmission with an extremely strong dispersion, which shows great potentials for applications in slow light devices [25,26], optical buffers [27,28], ultrasensitive biosensing [29], and enhanced nonlinear effects [30].

In this letter we report, for the first time to our knowledge, about a completely new mechanism that opens a transparency window in an opaque medium using an external magnetic field, i. e. magnetically induced transparency (MIT). We consider a photonic crystal layer based on a metamaterial with a helical periodical structure (cholesteric liquid crystal (CLC)-like structure) in an external magnetic field, which exhibits magneto-optical activity. The dielectric permittivity and magnetic permeability tensors have the forms:

$$\hat{\varepsilon}(z) = \varepsilon_m \begin{pmatrix} 1 + \delta\cos 2az & \pm\delta\sin 2az \pm ig/\varepsilon_m & 0 \\ \pm\delta\sin 2az \mp ig/\varepsilon_m & 1 - \delta\cos 2az & 0 \\ 0 & 0 & 1 - \delta \end{pmatrix}, \hat{\mu}(z) = \hat{I}, \qquad (1)$$

where $\varepsilon_m = (\varepsilon_1 + \varepsilon_2)/2$, $\delta = \frac{(\varepsilon_1 - \varepsilon_2)}{(\varepsilon_1 + \varepsilon_2)}$, $\varepsilon_{1,2}$ are the principal values of the local dielectric permittivity tensor, g is the parameter of magnetooptical activity, $a = 2\pi/p$, $p$ is the helix pitch. We consider the case of light propagation along the helix axis (along the z-axis). An exact analytical solution of the Maxwell equations for a magnetoactive CLC-like layer in the case of propagation of the light along its optical axis (helix axis) is known [31]. Using it, we can solve the problem of reflection, transmission, and localization of light in the case of a magnetoactive CLC-like layer of finite thickness. We assume that the optical axis of the CLC-like layer is perpendicular to the boundaries and directed along the z-axis. The CLC-like layer on both its sides border with isotropic half-spaces with the same refractive indices equal to $n_s$. The boundary conditions, consisting of the continuity of the tangential components of the electric and magnetic fields, are a system of eight linear equations with eight unknowns. Thus, solving this boundary-value problem, we can determine the values for the components of the reflected $\vec{E}_r(z)$ and transmitted $\vec{E}_t(z)$ fields, as well as for the field $\vec{E}_{in}(z)$ in the CLC layer itself and, therefore, calculate the reflection $R = |E_r|^2/|E_i|^2$, transmission $T = |E_t|^2/|E_i|^2$, and absorption $A = 1 - (R + T)$ coefficients. Further, below, all calculations were made for magnetoactive CLC with the following parameters $\varepsilon_1 = 0.8$, $\varepsilon_2 = 0.35$, helix pitch is $p = 400$ nm, layer thickness is $d = 5p$ and $n_s = \sqrt{\varepsilon_m}$. The choice of such parameters is dictated by the fact that the effects revealed below are more clearly demonstrated at these values of the medium parameters.

Fig. 1 shows (a,b) the absorption $A$ and (c,d) transmission $T$ spectra. The light incident on the layer has a non-diffracting (a,c) and diffracting eigen polarization (EP). The EPs are the two polarizations of the incident light, which do not change when light transmits through the system [32]. They connected with the polarizations of the excited in the medium internal waves (the eigenmodes). For CLCs with weak local birefringence, the two EPs practically coincide with right and left circular polarizations.

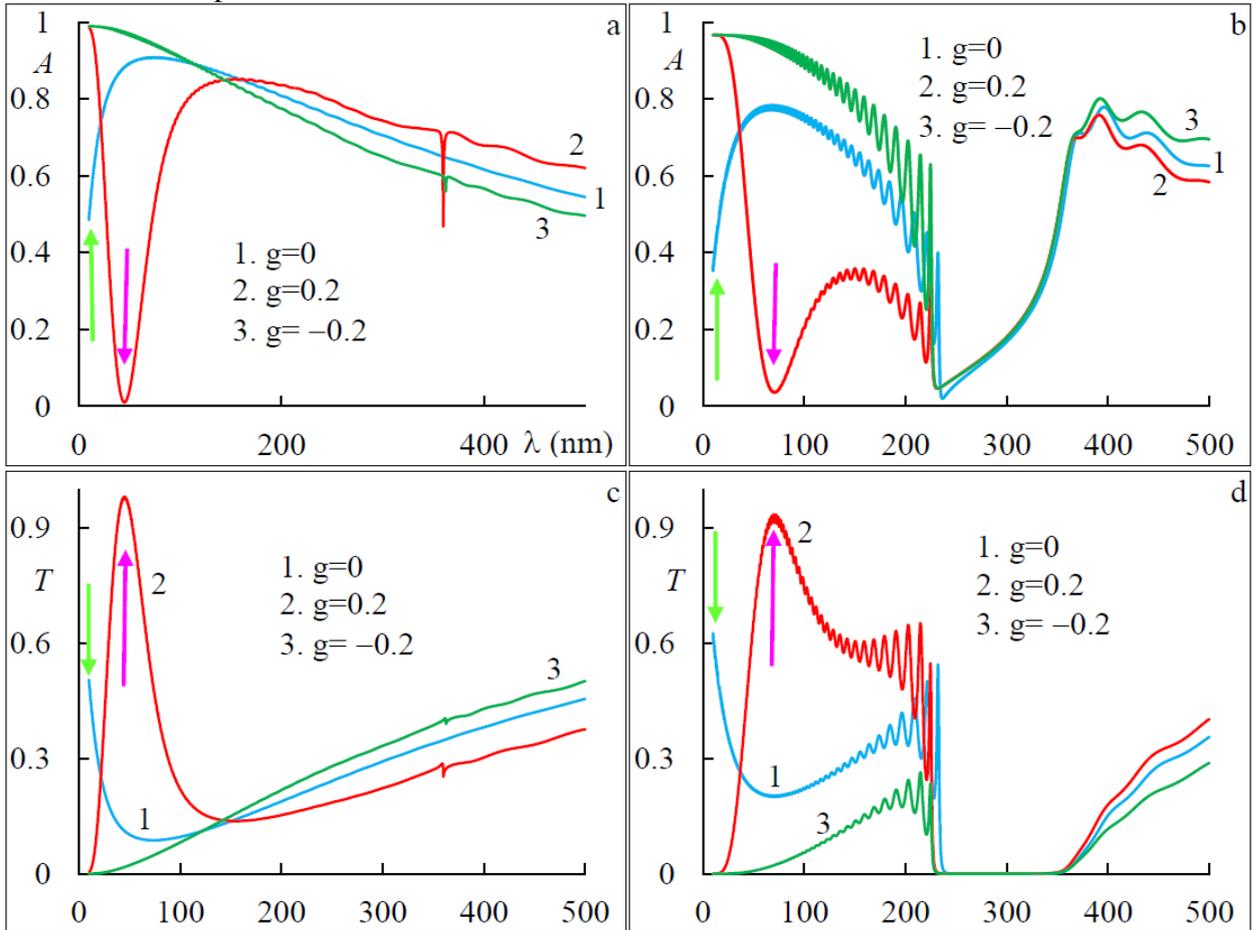

Fig. 1. (a,b) The absorption $A$ and (c,d) transmission $T$ spectra in the case of anisotropic absorption (a,c: $\mathrm{Im}\varepsilon_1 = 0.$, $\mathrm{Im}\varepsilon_2 = 0.05$; b, d: $\mathrm{Im}\varepsilon_1 = 0.05$, $\mathrm{Im}\varepsilon_2 = 0.$).

From the presented spectra, it follows that:
1) when $\mathrm{Im}\varepsilon_1 = 0.$, $\mathrm{Im}\varepsilon_2 = 0.05$ for incident light with a non-diffracting EP there is a transparency window, where $A \approx 0$ and $T \approx 1$;
2) for incident light with a diffracting EP a transparency window exists when $\mathrm{Im}\varepsilon_1 = 0.05$, $\mathrm{Im}\varepsilon_2 = 0.$ (in Fig.1 all these transparency windows are indicated by purple arrows);
3) these windows appear only in the presence of external magnetic field (at $g \neq 0$);
4) this effect is nonreciprocal, it takes place at $g = 0.2$, but does not take place at $g = -0.2$;
5) this effect does not take place at isotropic absorption, that is in the case $\mathrm{Im}\varepsilon_1 = \mathrm{Im}\varepsilon_2$;
6) at the absence of external magnetic field (at $g = 0.$) in the case of anisotropic absorption $T \to 1$ and $A \to 0$, when $\lambda \to 0$ (these points in the figure are indicated by green arrows).

So, we registered a new effect, namely MIT. And now, in order to identify the reasons for the revealed effect MIT, we firstly will investigate the features of the wave vectors of eigen modes exited in medium. As well-known the dispersion equation for magnetoactive CLC-like medium at light propagation along helix axis has the form [33]:

$$k_z^4 + a_1 k_z^2 + a_2 k_z + a_3 = 0, \quad m = 1,2,3,4, \tag{2}$$

where $a_1 = -2\left(\frac{\omega^2}{c^2}\varepsilon_m + a^2\right)$, $a_2 = -4\frac{\omega^2}{c^2}ag$, $a_3 = -2\frac{\omega^2}{c^2}a^2\varepsilon_m + \frac{\omega^4}{c^4}\varepsilon_m^2(1-\delta^2) - \frac{\omega^4}{c^4}g^2 + a^4$, $\omega$ is the frequency and $c$ is the light speed in vacuum. Two of these modes are diffracting (there exist a wavelength band where these wave vectors are complex at the absence of absorption) and two others are propagating modes (they are real at the absence of absorption). As showed in [33] the photonic band gap (PBG) takes blueshift in the external magnetic field. Furthermore, in the case g > 0, the real parts of the $k_m(\lambda)$ of propagating modes are displaced downwards, and of the diffracting modes are displaced upwards. In the case of g < 0 we have the opposite picture. As showed in our simulations in the presence of external magnetic field are displaced not only the real parts of wave vectors of diffracting and propagating modes but also their imaginary parts and which causes the observed effect to be fulfilled. Fig. 2 shows the dependences of Im$k(\lambda)$ for nondiffracting (a) and diffracting (b) eigen modes. As it is seen in Fig.2 at g ≠ 0 there appear regions with Im$k$=0 which are indicated by purple arrows, i.e., the external magnetic field makes the absorbing medium transparent. Further, as can be seen from this figure, for g = 0 Im$k \to 0$, when $\lambda \to 0$ (indicated in figure by green arrows), which explains the fact that $T \to 1$ and $A \to 0$, when $\lambda \to 0$. At isotropic absorption in this limit Im$k$ for propagating and diffracting eigen modes are rise indefinitely, moreover as at presence of external magnetic field as its absence.

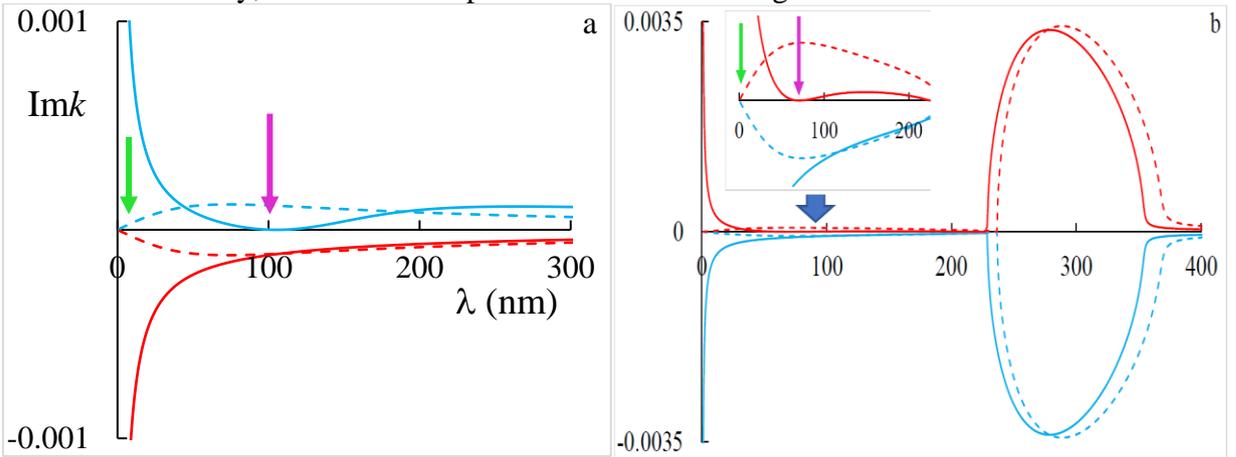

Fig.2. The dependences of Im$k(\lambda)$ for nondiffracting (a) and diffracting (b) eigen modes at the absence of external magnetic field (g=0, Im$\varepsilon_1 = 0.$, Im$\varepsilon_2 = 0.05$; dashed lines) and at its presence (g=0.2; Im$\varepsilon_1 = 0.05$, Im$\varepsilon_2 = 0.$, solid lines).

Now we pass to demonstrate the tunability of this effect. Fig. 3 shows the transmission $T$ spectra in the case Im$\varepsilon_1 = 0.$, Im$\varepsilon_2 \neq 0$, a: at different values of a parameter g, and b: at different values of a Im$\varepsilon_2$. The light incident on the layer has a non-diffracting EP. As follows from presented results by changing parameter g (external magnetic field) we can continuously change the wavelength of transparency band and by changing the parameter Im$\varepsilon_2$ we can change the frequency width of transparency line. Let us note that the same type dependencies are observed in the case Im$\varepsilon_2 = 0.$, Im$\varepsilon_1 \neq 0$ for incident light with diffracting EP.

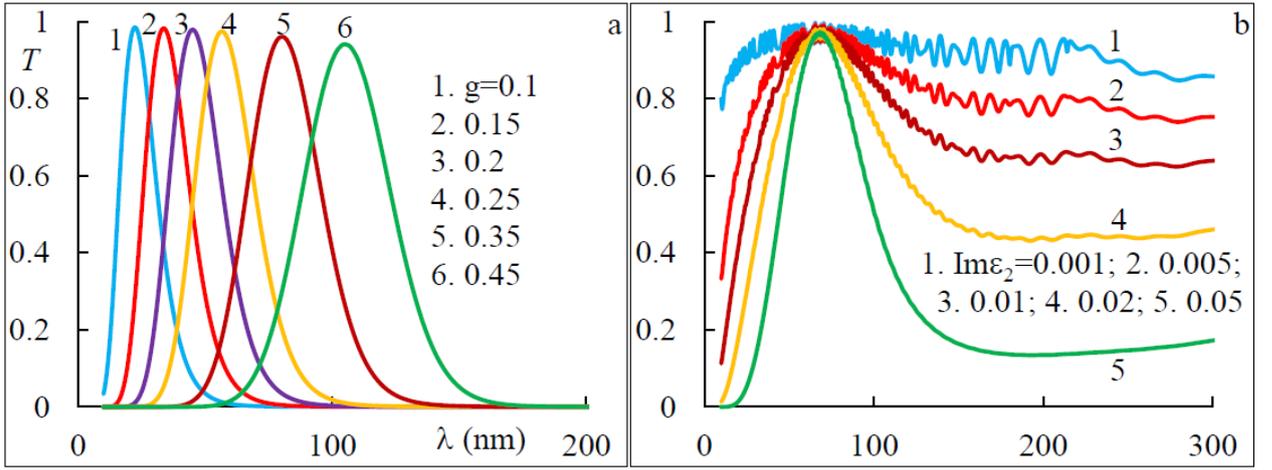

Fig. 3. The transmission $T$ spectra.

Up to now we did not consider the effects of optical dispersion or dispersion absorption, i. e. we supposed that the components of the dielectric and magnetic tensors are constant and do not depend on the frequency; the imaginary parts also do not depend on the frequency. Now we pass to more general case taking into account the dispersion dependences of optical parameters of CLC-like medium. We here assume the Lorentzian form of frequency dependence of dielectric tensor components, i.e. we assume that

$$\varepsilon_{1,2}(\omega) = \varepsilon_{1.2} + \frac{f_{1,2}}{\omega_{01,2}^2 - \omega^2 - i\gamma_{1,2}\omega}, \qquad (3)$$

where $\gamma_{1,2}$ are the broadenings of resonance absorption lines or simply damping factors, $f_{1,2}$ are the quantities proportional to oscillator strengths, $\omega_{01,2}$ are the resonance frequencies, and $\varepsilon_{1.2}$ are parts of dielectric permittivities not depending on frequency and $i$ is the imaginary unit. Fig. 4 shows (a) the $\mathrm{Im}\varepsilon_2$ (solid line) and $\mathrm{Re}\varepsilon_2$ (dashed line) and (b) the absorption $A$ (solid line) and transmission (dashed line) spectra. The light incident on the layer has a non-diffracting EP (b). As follows from presented results the external magnetic field makes a medium transparent in a narrow spectral range around the absorption line.

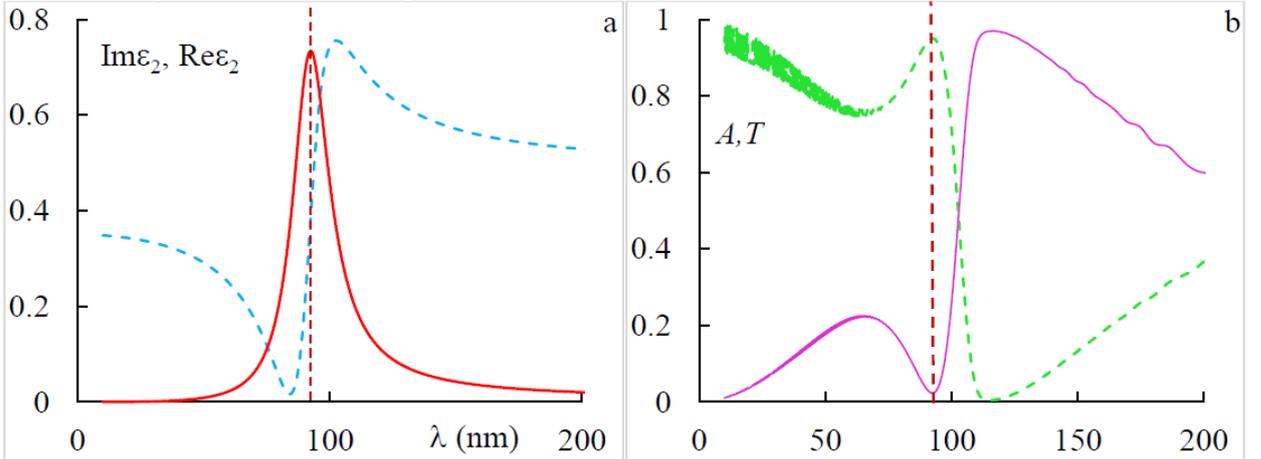

Fig. 4. (a) The $\mathrm{Im}\varepsilon_2$ (solid line) and $\mathrm{Re}\varepsilon_2$ (dashed line) and (b) the absorption $A$ (solid line) and transmission (dashed line) spectra. The rest of parameters are: g=0.4, $f_1 = 0$, $f_2 = 6 \cdot 10^{31}$ s$^{-2}$, $\gamma_2 = 4 \cdot 10^{15}$ s$^{-1}$, $\omega_{02} = 2\pi c/\lambda_{02}$, $\lambda_{02} = 92$ nm.

Let us note, that, as show our simulations, discovered effect takes place also for usual CLCs, for instance for CLC cholesteryl-nonanoate–cholesteryl-chloride–cholesteryl-acetate (composition 20:15:6), with parameters $\varepsilon_1 = 2.29$, $\varepsilon_2 = 2.143$ and which at the temperature $t = 25$ °C has helix pitch $p$=420 nm.

In conclusion we show theoretically the existence of a MIT in a magnetoactive CLC. In addition, the capability for creating such transparency window, and for controlling the linewidth of such resonance and transparency band frequency, is important for applications such as tunable bandwidth filter, and etc.

**References**
1. S. E. Harris, Phys. Today **50**, 36 (1997).
2. R. Röhlsberger, H.-C. Wille, K. Schlage, B. Sahoo, Nature **482**, 199–203 (2012).
3. M. Mücke, et al, Nature **465**, 755–758 (2010).
4. N. Papasimakis, et al., Phys. Rev. Lett. **101**, 253903 (2008).
5. A. Jain, et al., Phys. Rev. Lett. **109**, 187401 (2012).
6. R. Singh, C. Rockstuhl, F. Lederer, W. Zhang, Phys. Rev. B **79**, 085111 (2009).
7. P. Tassin, et al., Phys. Rev. Lett. **102**, 053901 (2009).
8. N. Liu, et al. Nat. Mater. **8**, 758 (2009).
9. S. Zhang, D.A. Genov, Y. Wang, M. Liu, X Zhang, Phys. Rev. Lett. **101**, 047401 (2008).
10. N. Verellen, et al., Nano Lett. **9**, 1663 (2009).
11. R. Taubert, M. Hentschel, J. Kästel, H. Giessen, Nano Lett. **12**, 1367 (2012).
12. G.C. Dyer, et al., Nat. Photon. **7**, 925 (2013).
13. C.W. Hsu, et al., Nano Lett. **14**, 2783 (2014).
14. Q. Xu, et al., Phys. Rev. Lett. **96**, 123901 (2006).
15. Q. Xu, P. Dong, M. Lipson, Nat. Phys. **3**, 406–410 (2007).
16. M.F. Limonov, M.V. Rybin, A.N. Poddubny, Y.S. Kivshar, Nat. Photon. **11**, 543 (2017).
17. C. Wang, et al., Nat. Phys. **16**, 334–340 (2020).
18. S. Weis, et al. Science **330**, 1515 (2010).
19. A. H. Safavi-Naeini, et al., Nature **472**, 69 (2011).
20. H. Lü, C. Wang, L. Yang, H. Jing, Phys. Rev. Appl. **10**, 14006 (2018).
21. P.M. Anisimov, J.P. Dowling, B.C. Sanders, Phys. Rev. Lett. **107**, 163604 (2011).
22. A.A. Abdumalikov, et al., Phys. Rev. Lett. **104**, 193601 (2010).
23. Q. Xu, et al., Phys. Rev. Lett. **96**, 123901 (2006).
24. X. Yang, M. Yu, D.-L. Kwong, C. Wong, Phys. Rev. Lett. **102**, 173902 (2009).
25. X. Yin, et al., Appl. Phys. Lett., **103**, 021115 (2013).
26. P. Tassin, et al., Phys. Rev. Lett., **102**, 053901 (2009).
27. W. Cao, et al., Appl. Phys. Lett., **103**, 101106 (2013).
28. R. Singh, I. A. et al, Appl. Phys. Lett., **99**, 201107 (2011).
29. C. Wu, A. B. Khanikaev, R. Adato, N. Arju, A. A. Yanik, H. Altug, G. Shvets, Nat. Mater., **11**, 69 (2012).
30. Y. Wu, J. Saldana, Y. Zhu, Phys. Rev. A, **67**, 013811 (2003).
31. A.H.Gevorgyan. Mol. Cryst. Liquid Cryst., **378,** 129 (2002).
32. R. M. A. Azzam and N. M. Bashara, Ellipsometry and polarized light (North-Holland, Amsterdam, 1977).
33. A.H. Gevorgyan. Optical Materials, **113**, 110807 (2021).